\documentclass[12pt]{article}
\usepackage{epsf}
\usepackage{amsmath}
\usepackage{graphicx}
\usepackage{cite}

\renewcommand{\thefootnote}{\#\arabic{footnote}}
\begin{document}

\newcommand{\gtrsim}{ \mathop{}_{\textstyle \sim}^{\textstyle >} }
\newcommand{\lesssim}{ \mathop{}_{\textstyle \sim}^{\textstyle <} }

\newcommand{\rem}[1]{{\bf #1}}

\renewcommand{\thefootnote}{\fnsymbol{footnote}}
\setcounter{footnote}{0}
\begin{titlepage}

\def\thefootnote{\fnsymbol{footnote}}

\begin{center}
{\Large \bf Black Holes as Dark Matter}

\vskip .5in

{\bf Paul H. Frampton\footnote{frampton@physics.unc.edu}

{Department of Physics and Astronomy, University of North Carolina at Chapel Hill}

CB3255, Chapel Hill, NC 27599-3255, USA}

\end{center}

\vskip .4in

\begin{abstract}
\noindent 
 While the energy of the universe has been established to be
 about
 0.04 baryons, 0.24 dark matter and 0.72 dark energy,
the cosmological entropy
 is almost entirely, about $(1 - 10^{-15})$, from black holes
 and only $10^{-15}$ from everything else.
This identification of all dark matter as black holes
is natural in statistical mechanics.
Cosmological history of dark matter is discussed.
\end{abstract}

\end{titlepage}

\newpage

\section{Introduction}

\noindent The present article discusses a solution
for the cosmological
dark matter problem.
The problem is over 75 years old\cite{Zwi33} and
has been the subject of innumerable papers, conferences and books.

\noindent There is about six times more dark matter than baryonic
matter in the universe. A well-known book\cite{KoTu}
on the connection between particle phenomenology
and theoretical cosmology discusses possible candidates
-- axions, WIMPs, MACHOs -- for the constituents of the dark matter. In the
present article I shall provide a complete solution
to the dark matter problem which is in the MACHO category
and surprisingly simple: the
dark mattter is all black holes!

\noindent Let me first consider low mass
dark matter
candidates suggested by particle phenomenology. 
There are two especially popular candidates. 

The lighter
is the invisible axion predicted 
by \cite{dfs81,ki79,svz80,zh80}. Its mass
\footnote{Throughout we adopt an order-of-magnitude
notation where $10^x$ denotes any integer
between $10^{x-1}$ and $10^{x+1}$.}
must be in
an allowed window $10^{-15} GeV < M_{axion} < 10^{-12} GeV$.
Although there are compelling arguments\cite{ho92}
that the invisible axion requires stronger fine tuning
than the strong CP problem it purports to solve,
it has staying power and remains the subject of experimental
searches.

\noindent The other very popular
low-mass dark matter candidate,
at a proposed mass typcally a trillion times
that of the invisible axion, is 
the WIMP (Weakly-Interacting Massive Particle)
exemplified by the neutralino of supersymmetry. The
fact that WIMPs can naturally annihilate to an acceptable
present abundance of dark matter is sometimes cited
as support for WIMPs as dark matter.

\noindent But if my proposal for identification of the constituents
of dark matter as black holes with
masses above $30 M_{\odot}$ is correct, particle-phenomenology-inspired
dark matter candidates 
are, unfortunately for their creators, irrelevant.

\noindent Higher mass dark matter constituents are generically 
called MACHOs (Massive Astrophysical
Compact Halo Objects). My proposed constituents fall into this class
with masses in the range $30 M_{\odot} <M_{MACHO} < 500 M_{\odot}$
where $M_{\odot}$ is the solar mass; equivalently 
$10^{26} kg < M_{MACHO} < 10^{27} kg$ or
$10^{55} GeV < M_{MACHO} < 10^{56}$ GeV.
Even the most die-hard opponent to the Large Hadron
Collider (LHC) could not take seriously that the LHC
could create a black hole with fifty orders of magnitude
times the available LHC energy of $10^4 GeV$ so the present article thus
might convince that the LHC is safe.

\noindent The MACHOs are truly compact, all being smaller
in physical size than the Earth.
It is the small size and large mass which have
enabled these dark matter constituents
to escape detection for 75 years. We discuss two
methods for their detection, their formation 
then what, for me, is the best motivation - the entropy of the universe.

\bigskip

\newpage

\begin{figure}
\begin{center}
\vspace{15pt}
\includegraphics[height=130mm]{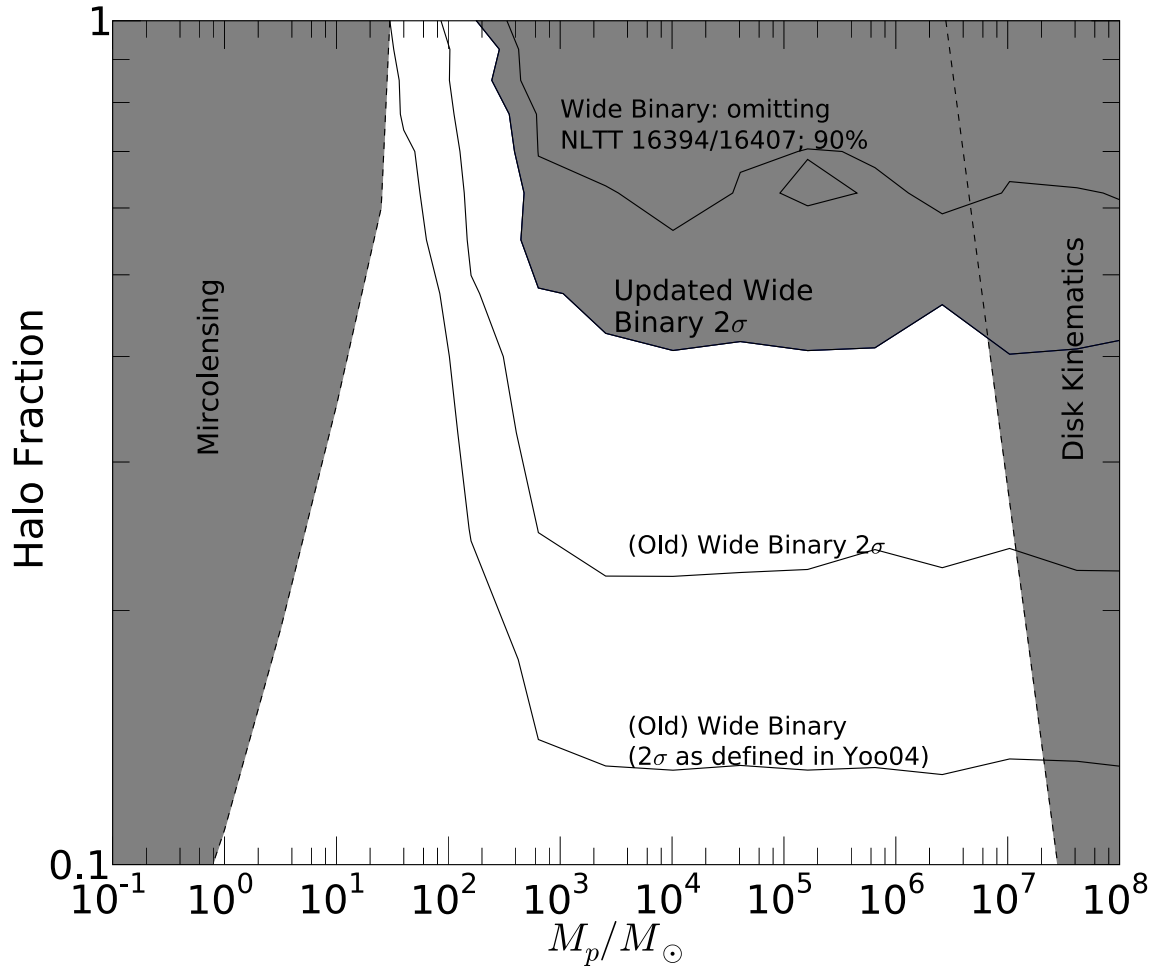}
\vspace{15pt}
\caption{\normalsize 
Fig. 1. The vertical axis is the total intermediate-mass 
black hole (IMBH)
mass as the percent of the halo dark matter; the
horizontal axis is the individual IMBH mass in terms of the solar
mass $M_{\odot}$.
Note that for the range of masses between $30M_{\odot}$ and
$500 M_{\odot}$ all
halo mass can be IMBHs.
Taken from \cite{Qui09}.}
\label{Figure}
\end{center}
\end{figure}

\bigskip

\newpage

\section{MACHO Searches by Microlensing}

\noindent In the late twentieth century heroic
searches were made for microlensing events and
many spectacular examples \cite{al97,al00}
of MACHOs were identified
with masses up to $30M_{\odot}$. The length of time
of a microlensing event depends on the velocity
and mass of the lensing MACHO. Since the velocities
do nor vary widely it is the MACHO mass which
is most crucial in determining longevity.

\noindent The longevity is directly proportional
to the square root of the lens mass. MACHOS
were discovered with microlensing longevities
up to about 400 days. Insufficient MACHOs to
account for more than a few percent
of the dark matter were discovered.
To detect MACHOs up to
mass $500 M_{\odot}$, the current upper limit
for halo black holes, necessitates 
study of microlensing events with duration up to
4.5 years.

\section{Wide Binaries}

\noindent Nature has provided delicate astrophysical clocks
whose accuracy is sensitive to the possible
proximity of MACHOs. These clocks are binaries
with separations up to 1 pc and therefore
can be bigger than the Solar System in which the
distance from the Sun to the outermost
planet is only $10^{-4}$ pc.

\noindent Because these wide binaries are so weakly bound
they are the most sensitive to disruption by nearby
MACHOs. After formation they retain their original orbital
parameters as a reliable clock except when affected 
by gravitational encounters with MACHOs.

\noindent The detailed study of wide binaries is a young
and promising technique for MACHO detection.
In the first pioneering presentation\cite{Yoo04} the range
of masses allowed for MACHOs which provide all
dark matter was $30M_{\odot} < M_{MACHO} < 43M_{\odot}$.
It seemed unlikely, though logically possible, that
all dark matter MACHOs could 
have masses within such a narrow range.

\noindent A revisit and reanalysis\cite{Qui09} of
wide binaries has, however, recently
\footnote{I am grateful to Julio Chaname
for bringing \cite{Qui09}
to my attention.}
discovered that one of the key examples used in
\cite{Yoo04} was spurious. This leads to quite
different and much more promising
restrictions on MACHOs, with the allowed mass
range extending up to $500 M_{\odot}$.

\noindent The new constraint 
is displayed in Fig. 1. and it is now
far more likely that all dark matter 
is in the form of black hole MACHOs.  

\section{Formation} 

\noindent Because intergalactic dust shows evidence of metallicity
which involves nuclei heavier than lithium, the heaviest element
formed during big-bang nucleosynthsis, it has been hypothesized
\cite{mr01} that there must have existed Pop III stars 
in an era before galaxy formation.
These subsequently exploded producing the dust nuclei heavier
than lithium, and leaving massive black holes which
are MACHO candidates.

\noindent Numerical simulations of dark matter halos\cite{nfw96}
have given insight into the likely general features
of the halo profile. The spatial
resolution of such simulations is, however,
far too coarse to
study the formation of MACHOs smaller than the Earth
whose radius is $10^{-10} pc$.
Nevertheless, dark matter is known to clump
at large scales and may well do so at such 
far smaller scales.

\section{Irrelevance of Dark Energy}

\noindent I have assumed that dark energy possesses no
entropy. Let me argue that this is a very weak assumption.

\noindent Recently the Planck satellite was
launched and is on its way to Lagrangian point 2
where it will attempt to measure with unprecedented
accuracy the dark energy
equation of state $w = p/\rho$ where $p, \rho$ are
respectively pressure, density. 

\noindent The key quantity is $\phi = (1 + w)$.
If $\phi$ vanishes identically the assumption
of zero entropy for dark energy is justified because
it is fully described by one parameter, the
cosmological constant.
If non-zero $\phi$ were observationally established,
it would imply a dynamical dark energy with
concomitant entropy. However, without black holes,
the dark energy dimensionless
entropy would be much less than one googol
and so the identification of the dark matter
as black holes and the rest of the
present discussion would be unchanged.

\section{Discussion}

\noindent The identification of the dark matter constituent
has been an intellectual pursuit for over 75 years.
Particle theorists have naturally suggested light mass
solutions which include the invisible axion and WIMPs.
According to theoretical cosmology, however, the most
likely solution is that all dark matter is black holes.
According to the recently updated
analysis of wide binaries the
permitted mass range is between 30 and 500 solar masses.

\noindent The present dimensionless entropy of the universe
is 1000 googols. This is useful for
model building in cyclic cosmology
as an alternative to the big bang
\cite{bf07,bf08,bfm08,pf07,pf09}.

\noindent Taking a universe comprised
of $10^{11}$ halos each of mass $10^{12} M_{\odot}$,
and using a central MACHO mass $100 M_{\odot}$,
there are $10^{10}$ MACHOs per halo and $10^{21}$
in the universe. The dimensionless entropy
per MACHO, using the PBH formula
\cite{pa69,be73,ha74}
is $10^{82}$ and the total is 1000 googols.
For galactic-core supermassive black holes (SMBHs), there
are $10^{11}$ of them so that for a central mass
$10^{7} M_{\odot}$ and individual entropy $10^{92}$ their
total entropy is also 1000 googols.

\noindent Because $S \propto M^2$, the total mass of MACHOs
is much larger than that of SMBHs. For the central values
it is $10^5$ times larger. The total MACHO mass
is $10^{23} M_{\odot} = 10^{53} kg = 10^{80} GeV$
so I hereby
predict several thousand trillion trillion trillion trillion
kilograms of black holes!

\noindent The existence of such numerous black holes
will hopefully be confirmed by observational
astronomers. The two most promising known techniques
are microlensing and study of wide binaries.

\begin{center}

{\bf Acknowledgement}

\end{center}

\noindent
This work was supported by U.S. Department of Energy 
grant number DE-FG02-06ER41418.

\end{document}